\documentclass[aps,prd,twocolumn,showpacs,superscriptaddress,groupedaddress,floatfix]{revtex4}  
\usepackage{graphicx}  
\usepackage{dcolumn}   
\usepackage{bm}        
\usepackage{amssymb}   
\usepackage{color} 
\usepackage{xspace} 
\usepackage{amsmath}

\newcommand{\sigeff}{$\sigma_{\rm eff}$}
\newcommand{\sigteff}{$\sigma_{\rm eff}$}
\newcommand{\ptjs}{$p^{J/\psi}_T$}
\newcommand{\etajs}{$|\eta^{J/\psi}|$}
\newcommand{\DJP}{DJ~}

\begin{document}

\hspace{5.2in} \mbox{FERMILAB-PUB-14-176-E}

\title{Observation and studies of  double ${\bm {J/\psi}}$ production at the Tevatron}

%
\affiliation{LAFEX, Centro Brasileiro de Pesquisas F\'{i}sicas, Rio de Janeiro, Brazil}
\affiliation{Universidade do Estado do Rio de Janeiro, Rio de Janeiro, Brazil}
\affiliation{Universidade Federal do ABC, Santo Andr\'e, Brazil}
\affiliation{University of Science and Technology of China, Hefei, People's Republic of China}
\affiliation{Universidad de los Andes, Bogot\'a, Colombia}
\affiliation{Charles University, Faculty of Mathematics and Physics, Center for Particle Physics, Prague, Czech Republic}
\affiliation{Czech Technical University in Prague, Prague, Czech Republic}
\affiliation{Institute of Physics, Academy of Sciences of the Czech Republic, Prague, Czech Republic}
\affiliation{Universidad San Francisco de Quito, Quito, Ecuador}
\affiliation{LPC, Universit\'e Blaise Pascal, CNRS/IN2P3, Clermont, France}
\affiliation{LPSC, Universit\'e Joseph Fourier Grenoble 1, CNRS/IN2P3, Institut National Polytechnique de Grenoble, Grenoble, France}
\affiliation{CPPM, Aix-Marseille Universit\'e, CNRS/IN2P3, Marseille, France}
\affiliation{LAL, Universit\'e Paris-Sud, CNRS/IN2P3, Orsay, France}
\affiliation{LPNHE, Universit\'es Paris VI and VII, CNRS/IN2P3, Paris, France}
\affiliation{CEA, Irfu, SPP, Saclay, France}
\affiliation{IPHC, Universit\'e de Strasbourg, CNRS/IN2P3, Strasbourg, France}
\affiliation{IPNL, Universit\'e Lyon 1, CNRS/IN2P3, Villeurbanne, France and Universit\'e de Lyon, Lyon, France}
\affiliation{III. Physikalisches Institut A, RWTH Aachen University, Aachen, Germany}
\affiliation{Physikalisches Institut, Universit\"at Freiburg, Freiburg, Germany}
\affiliation{II. Physikalisches Institut, Georg-August-Universit\"at G\"ottingen, G\"ottingen, Germany}
\affiliation{Institut f\"ur Physik, Universit\"at Mainz, Mainz, Germany}
\affiliation{Ludwig-Maximilians-Universit\"at M\"unchen, M\"unchen, Germany}
\affiliation{Panjab University, Chandigarh, India}
\affiliation{Delhi University, Delhi, India}
\affiliation{Tata Institute of Fundamental Research, Mumbai, India}
\affiliation{University College Dublin, Dublin, Ireland}
\affiliation{Korea Detector Laboratory, Korea University, Seoul, Korea}
\affiliation{CINVESTAV, Mexico City, Mexico}
\affiliation{Nikhef, Science Park, Amsterdam, the Netherlands}
\affiliation{Radboud University Nijmegen, Nijmegen, the Netherlands}
\affiliation{Joint Institute for Nuclear Research, Dubna, Russia}
\affiliation{Institute for Theoretical and Experimental Physics, Moscow, Russia}
\affiliation{Moscow State University, Moscow, Russia}
\affiliation{Institute for High Energy Physics, Protvino, Russia}
\affiliation{Petersburg Nuclear Physics Institute, St. Petersburg, Russia}
\affiliation{Instituci\'{o} Catalana de Recerca i Estudis Avan\c{c}ats (ICREA) and Institut de F\'{i}sica d'Altes Energies (IFAE), Barcelona, Spain}
\affiliation{Uppsala University, Uppsala, Sweden}
\affiliation{Taras Shevchenko National University of Kyiv, Kiev, Ukraine}
\affiliation{Lancaster University, Lancaster LA1 4YB, United Kingdom}
\affiliation{Imperial College London, London SW7 2AZ, United Kingdom}
\affiliation{The University of Manchester, Manchester M13 9PL, United Kingdom}
\affiliation{University of Arizona, Tucson, Arizona 85721, USA}
\affiliation{University of California Riverside, Riverside, California 92521, USA}
\affiliation{Florida State University, Tallahassee, Florida 32306, USA}
\affiliation{Fermi National Accelerator Laboratory, Batavia, Illinois 60510, USA}
\affiliation{University of Illinois at Chicago, Chicago, Illinois 60607, USA}
\affiliation{Northern Illinois University, DeKalb, Illinois 60115, USA}
\affiliation{Northwestern University, Evanston, Illinois 60208, USA}
\affiliation{Indiana University, Bloomington, Indiana 47405, USA}
\affiliation{Purdue University Calumet, Hammond, Indiana 46323, USA}
\affiliation{University of Notre Dame, Notre Dame, Indiana 46556, USA}
\affiliation{Iowa State University, Ames, Iowa 50011, USA}
\affiliation{University of Kansas, Lawrence, Kansas 66045, USA}
\affiliation{Louisiana Tech University, Ruston, Louisiana 71272, USA}
\affiliation{Northeastern University, Boston, Massachusetts 02115, USA}
\affiliation{University of Michigan, Ann Arbor, Michigan 48109, USA}
\affiliation{Michigan State University, East Lansing, Michigan 48824, USA}
\affiliation{University of Mississippi, University, Mississippi 38677, USA}
\affiliation{University of Nebraska, Lincoln, Nebraska 68588, USA}
\affiliation{Rutgers University, Piscataway, New Jersey 08855, USA}
\affiliation{Princeton University, Princeton, New Jersey 08544, USA}
\affiliation{State University of New York, Buffalo, New York 14260, USA}
\affiliation{University of Rochester, Rochester, New York 14627, USA}
\affiliation{State University of New York, Stony Brook, New York 11794, USA}
\affiliation{Brookhaven National Laboratory, Upton, New York 11973, USA}
\affiliation{Langston University, Langston, Oklahoma 73050, USA}
\affiliation{University of Oklahoma, Norman, Oklahoma 73019, USA}
\affiliation{Oklahoma State University, Stillwater, Oklahoma 74078, USA}
\affiliation{Brown University, Providence, Rhode Island 02912, USA}
\affiliation{University of Texas, Arlington, Texas 76019, USA}
\affiliation{Southern Methodist University, Dallas, Texas 75275, USA}
\affiliation{Rice University, Houston, Texas 77005, USA}
\affiliation{University of Virginia, Charlottesville, Virginia 22904, USA}
\affiliation{University of Washington, Seattle, Washington 98195, USA}
\author{V.M.~Abazov} \affiliation{Joint Institute for Nuclear Research, Dubna, Russia}
\author{B.~Abbott} \affiliation{University of Oklahoma, Norman, Oklahoma 73019, USA}
\author{B.S.~Acharya} \affiliation{Tata Institute of Fundamental Research, Mumbai, India}
\author{M.~Adams} \affiliation{University of Illinois at Chicago, Chicago, Illinois 60607, USA}
\author{T.~Adams} \affiliation{Florida State University, Tallahassee, Florida 32306, USA}
\author{J.P.~Agnew} \affiliation{The University of Manchester, Manchester M13 9PL, United Kingdom}
\author{G.D.~Alexeev} \affiliation{Joint Institute for Nuclear Research, Dubna, Russia}
\author{G.~Alkhazov} \affiliation{Petersburg Nuclear Physics Institute, St. Petersburg, Russia}
\author{A.~Alton$^{a}$} \affiliation{University of Michigan, Ann Arbor, Michigan 48109, USA}
\author{A.~Askew} \affiliation{Florida State University, Tallahassee, Florida 32306, USA}
% ...
%
%
\collaboration{The D0 Collaboration\footnote{with visitors from
$^{a}$Augustana College, Sioux Falls, SD, USA,
$^{b}$The University of Liverpool, Liverpool, UK,
$^{c}$DESY, Hamburg, Germany,
$^{d}$Universidad Michoacana de San Nicolas de Hidalgo, Morelia, Mexico
$^{e}$SLAC, Menlo Park, CA, USA,
$^{f}$University College London, London, UK,
$^{g}$Centro de Investigacion en Computacion - IPN, Mexico City, Mexico,
$^{h}$Universidade Estadual Paulista, S\~ao Paulo, Brazil,
$^{i}$Karlsruher Institut f\"ur Technologie (KIT) - Steinbuch Centre for Computing (SCC),
D-76128 Karlsruhe, Germany,
$^{j}$Office of Science, U.S. Department of Energy, Washington, D.C. 20585, USA,
$^{k}$American Association for the Advancement of Science, Washington, D.C. 20005, USA,
$^{l}$Kiev Institute for Nuclear Research, Kiev, Ukraine,
$^{m}$University of Maryland, College Park, Maryland 20742, USA
and
$^{n}$European Orgnaization for Nuclear Research (CERN), Geneva, Switzerland
}} \noaffiliation
\vskip 0.25cm
\date{\today}

\begin{abstract}
We present the observation of doubly-produced $J/\psi$ mesons with the D0 detector at Fermilab in
$p\bar{p}$ collisions at $\sqrt{s}=1.96$ TeV.   The production cross section for both singly and
doubly-produced $J/\psi$ mesons is measured using a sample with an integrated luminosity of 8.1~fb$^{-1}$.
For the first time, the double $J/\psi$ production cross section is separated into contributions due to
single and double parton scatterings.
 Using these measurements, we determine the effective cross section \sigteff,
a parameter characterizing an effective spatial area of the parton-parton interactions and related  to the parton spatial density inside the nucleon.
\end{abstract}



\maketitle

Heavy quarkonium is a well established probe of both quantum chromodynamics (QCD) and possible new bound states of hadronic matter,
e.g., tetraquarks~\cite{JJ_LHC,humpert}. Production of multiple quarkonium states provides
 insight into the  parton structure of the nucleon and parton-to-hadron fragmentation effects. 
In $p\bar{p}$ collisions, there are three main production mechanisms for $J/\psi$ mesons:  
prompt production (i.e. directly at the interaction point)
of $J/\psi$, and prompt production of heavier charmonium states, such as the 
$^3{}\!P_1$ state $\chi_{1c}$ and the $^3{}\!P_2$ state $\chi_{2c}$ that decay to $J/\psi+\gamma$, 
or decay to $J/\psi+X$ of directly produced $\psi(2S)$, and non-prompt B hadron decays. 
The first observation of  $J/\psi$ meson pair production was made in 1982 by the NA3 
Collaboration~\cite{NA3_1,NA3_2}.
The LHCb Collaboration has measured the  double $J/\psi$ production cross section 
in proton-proton collisions at $\sqrt{s}=7$ TeV~\cite{LHCb}.
At   Tevatron  and  LHC energies this cross section is  dominated by gluon fusion,
$gg \rightarrow J/\psi J/\psi$~\cite{JJ_LHC,barsnigzotnew}.

The interest in this channel originates  from the different mechanisms that can generate  simultaneous double $J/\psi$ (DJ) meson  production 
in single parton (SP) and double parton (DP) scatterings in a single hadron-hadron collision.
 A number of discussions of early experimental results~\cite{yuan,JJ_Tev} and more recent LHCb results~\cite{kom, barsnigzotnew},
show that the fraction of DP events at the Tevatron and especially at the LHC can be quite substantial. 
Since the initial state is dominated by $gg$ scattering,
the fraction of DP scatterings representing simultaneous, independent parton interactions, should  significantly depend on the spatial distribution
of gluons in a proton~\cite{Dok}.
 Other DP studies involving vector bosons and jets  probe the spatial distributions in processes with quark-quark or quark-gluon initial states~\cite{AFS,cdf1,cdf2,D0_2,atlas,cms_w2j}.
The measurement of the SP production cross section provides unique information
to constrain  parametrizations of the gluon parton distribution function (PDF) at low parton momentum fraction and energy scale,
where the gluon PDF has large uncertainty~\cite{GluonPDF}.
The production of $J/\psi$ mesons may proceed via two modes, color singlet 
and color octet~\cite{JJ_LHC, JJ_Tev, CDF, NRQCD}.
Predictions carried out using non-relativistic QCD (NRQCD)
show that the color singlet process in SP scattering contributes $\approx 90\%$ for the region of transverse momenta, $p^{J/\psi}_T \geq 4$ GeV/c, 
relevant for this measurement~\cite{JJ_Tev,NRQCD}.

In this Letter, we present first observation of  double $J/\psi$ production at the Tevatron and 
measurements of single and double $J/\psi$ production cross sections. 
For the first time, the latter is split into measurements of the SP and DP production cross sections. 
This allows us to extract the effective cross section~($\sigma_{\rm{eff}}$), 
a parameter related to an initial state parton spatial density
distribution within a nucleon (see, e.g., \cite{barsnigzotnew}):
\begin{equation}
\sigma_{\rm{eff}} = \frac{1}{2} \frac{\sigma(J/\psi)^2}{\sigma_{\mathrm {DP}}(J/\psi J/\psi)}.
\label{eq:s_eff}
\end{equation}
The factor of $1/2$ corresponds to the two indistinguishable
processes of single $J/\psi$ production \cite{Sjost, Threl}.

The measurements are based on the data sample collected 
by the D0 experiment at the Tevatron in proton-antiproton ($p\bar{p}$) collisions at the center-of-mass energy $\sqrt{s}=1.96~\rm{GeV}$, 
and corresponds to an integrated luminosity of $8.1\pm0.5$~fb$^{-1}$~\cite{d0lum}.

All cross section measurements are performed for prompt $J/\psi$ mesons 
with \ptjs$>4$ GeV/c and \etajs$<2$, where $\eta^{J/\psi}$ ~is the $J/\psi$ pseudorapidity~\cite{coord}. 
The $J/\psi$ mesons are fully reconstructed via their decay $J/\psi \rightarrow \mu^+ \mu^-$.
 The muons are required to have  transverse momenta $p^\mu_T>2$ GeV/c if their absolute pseudorapidities are $|\eta^\mu|<1.35$ or total momenta $|p^\mu|>4$ GeV/c if $1.35<|\eta^\mu|<2$.
The cross sections measured with these kinematic requirements are refered below as fiducial cross sections.

The D0 detector 
is a general purpose detector described in detail elsewhere~\cite{D0det}. 
The sub-detectors used in this analysis to select events at the trigger level and to reconstruct muons are the muon and the central tracking systems.
The central tracking system, used to reconstruct charged particle tracks, consists of the silicon microstrip tracker (SMT)~\cite{smt}
and a central fiber tracker (CFT) detector both placed inside a 1.9~T solenoidal magnet. The solenoidal magnet is located  inside  the central calorimeter, 
which is surrounded by the muon detector~\cite{RunI_muon}. 
The muon detector consists of three layers of drift tubes and three layers of plastic scintillators, one inside 1.9~T toroidal magnets and
two outside. The luminosity of colliding beams is measured using plastic scintillator arrays installed in front of the two end calorimeter cryostats~\cite{d0lum}.

Muons are identified as having either hits in all three layers of the muon detector or just in one layer 
in front of the toroids~\cite{RunII_muon}. They are also required to be
matched to a track reconstructed by the central tracking system as having at least one hit in the SMT and at least two hits in the CFT detectors.
The muon candidates must satisfy timing requirements to suppress cosmic rays.
 Their distance of closest approach to the beam line
has to be less than $0.5$ cm and their matching tracks have to pass within 2 cm 
along the beam ($z$) axis of the event interaction vertex. 
The $p\bar{p}$ interaction vertex should be within 60 cm of the center of the detector along beam axis. 
Events that have two such muons with opposite electric charge  that satisfy an invariant
mass requirement of $2.85 < M_{\mu \mu} < 3.35$ GeV are identified as single $J/\psi$ candidates. Events having two such pairs of muons 
are identified as \DJP candidates.
Background events are mainly due to random combinations of muons from $\pi^{\pm}$, $K^{\pm}$ decays, 
continuous non-resonant $\mu^{+}\mu^{-}$ production in   Drell-Yan events (both called ``accidental background''), 
and  B hadron  decays into a $J/\psi+X$. In the case of the DJ production,
the background may also be caused by associated production of $J/\psi$ meson and 
a muon pair not produced by a $J/\psi$ decay (``$J2\mu$'' events).

To properly normalize the cross section measurements
and to reduce the  backgrounds, we require  events to pass at least one of the  low-$p_T$ di-muon triggers. 
The single $J/\psi$  trigger efficiency is estimated using events which pass 
zero-bias triggers (which only require a beam crossing) or minimum bias triggers  
(which  only require  hits in the luminosity detectors), and  that also pass the  di-muon trigger.
The efficiency of the kinematic selections of the muons and $J/\psi$ mesons is found to be
 $0.124 \pm 0.024\thinspace\mbox{(stat)}  \pm 0.012\thinspace\mbox{(syst)} $. 
The systematic uncertainty is due to variations in the paramaterizations
of the functional forms used to fit the signal and background events to data.

To measure the trigger efficiency for double $J/\psi$ selection, we use DP and SP events generated in Monte Carlo (MC).
The double $J/\psi$ DP events are generated with the {\sc pythia}~\cite{pythia} MC event generator, while the double $J/\psi$ SP events are generated with {\sc herwig++}~\cite{herwig}.
 Events passed  through a {\sc geant} based~\cite{geant} simulation of the D0 detector and  overlaid  
with data  zero-bias events  are then processed with the same reconstruction code  as data. 
Using the di-muon trigger efficiency parametrized as a  2D function of the $p_T$ of each of the muons,
we calculate it for every possible pairing of muons in DP and SP MC events, 
and obtain efficiencies of $\varepsilon^{\rm{DP}}_{\rm{tr}}=0.48 \pm 0.07$ and $\varepsilon^{\rm{SP}}_{\rm{tr}}=0.51 \pm 0.07$,
where the uncertainty is propagated from the uncertainty on the di-muon trigger efficiency described above.

The number of single $J/\psi$ events after selections is about $7.4\times 10^6$. 
The  background from $\pi^{\pm}$, $K^{\pm}$ decays and DY events, in our single $J/\psi$ selection is estimated as a function of \ptjs and $\eta^{J/\psi}$.
In each (\ptjs, $\eta^{J/\psi}$) bin, we perform a simultaneous fit of signal using a double Gaussian function and background with a linear 
mass dependence in a window of $2.3 < M_{\mu \mu} < 4.2$ GeV.
We then  calculate the  background in the selection mass window of $2.85 < M_{\mu \mu} < 3.35$ GeV.
Averaging the contributions over all (\ptjs, $\eta^{J/\psi}$) bins, we estimate the  background fraction to be  $0.126 \pm 0.013$. 
The uncertainty is derived from variation of the fit parameters in the signal and background models.

We use {\sc pythia}  generated single $J/\psi$ events to estimate  the combined geometric and kinematic acceptance and 
reconstruction efficiency of the selection criteria, calculated as the ratio of the number of reconstructed events to the number of input events. 
The  generated events  are selected at the particle and reconstruction levels using the fiducial $J/\psi$ and muon 
kinematic selection criteria described above.
The number of reconstructed events is corrected for the different reconstruction efficiency in data and MC, calculated
in (\ptjs, $\eta^{J/\psi}$) bins.
The product of the acceptance and efficiency for 
single $J/\psi$ events produced in the color singlet model is found to be $0.221 \pm 0.002({\rm stat})\pm 0.023({\rm syst})$. 
The systematic uncertainty is  due to differences in the kinematic distributions 
between the simulated and data $J/\psi$ events, muon identification efficiency mismodeling, and differences between the color singlet and color octet models. 
The $\cos \theta^*$ distribution, 
where $\theta^*$ is the polar angle of the decay muon in the Collins-Soper frame~\cite{CS}, is sensitive to the $J/\psi$ polarization~\cite{baranov_polar,CDF_pol,CMS_pol}.  
Small data-to-MC reweighting factors based on the observed $\cos \theta^*$ are used to re-calculate the acceptance,
and lead to $\lesssim1\%$ difference  with the default acceptance value. 

Due to the long lifetimes of B hadrons, their decay vertex 
into the $J/\psi+X$ final state is usually several hundred microns away from 
the $p\bar{p}$ interaction  vertex, while prompt $J/\psi$ production occurs directly at
 the  interaction point. 
To distinguish prompt from non-prompt $J/\psi$ mesons,  
we examine  the decay length from the primary $p\bar{p}$ interaction vertex   to the $J/\psi$ production vertex,
defined as $c \tau = L_{xy} m^{J/\psi}_{\rm{pdg}}/p^{J/\psi}_T$, where $L_{xy}$ is the decay length of $J/\psi$ meson
calculated as the distance between the intersection of the muon tracks and the hard scattering vertex in the plane transverse to the beam, and $m^{J/\psi}_{\rm{pdg}}$
is the world average $J/\psi$ mass~\cite{PDG}.

\begin{figure}[h!]
~\\[-5mm]
\includegraphics[width=0.47\textwidth,keepaspectratio=true]{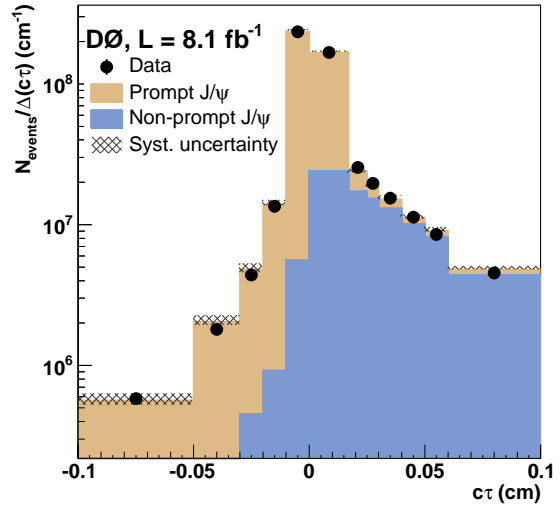}
~\\[-5mm]
\caption{(color online) 
The $c \tau$ distribution of background subtracted single $J/\psi$ events
after all selection criteria. The distributions for the
signal and background templates are shown normalized to
their respective fitted fractions. 
The uncertainty band corresponds to the total systematic uncertainty on the sum of signal and background events. 
}
\label{fig:single_jp}
\end{figure}

To estimate the fraction of prompt $J/\psi$ mesons in the data sample, we perform a maximum likelihood fit 
of the $c \tau$ distribution using templates 
for the prompt $J/\psi$ signal events,
taken from the single $J/\psi$ MC sample, and for non-prompt $J/\psi$ events,
taken from  the $b \bar b$ MC sample.
The latter are generated with {\sc pythia}~\cite{pythia}.
The prompt $J/\psi$ fraction obtained from the fit is $0.814 \pm 0.009$. The fit result is shown in Fig.~\ref{fig:single_jp}.
The overall $\chi^2/$ndf for the data/MC agreement for this fit varies, depending on the chosen SP and DP models,
within $0.50-0.85$ with corresponding $p$-values of $0.51-0.77$.
We verify that the \ptjs spectra of the  prompt signal (non-prompt background) events in data are well described by MC
in the signal (background) dominated regions by applying the selection $c \tau<0.02~(>0.03)$ cm.

The fiducial cross section of the prompt single $J/\psi$ production is calculated 
using the number of $J/\psi$ candidates in data, the fraction of prompt events, the dimuon trigger efficiency,
the acceptance and selection efficiency, as well as the integrated luminosity.
It is found to be
\begin{eqnarray}
\sigma(J/\psi) = 23.9 \pm 4.6 \mbox{(stat)} \pm 3.7 \mbox{(syst)} ~\mbox{nb}.
 \label{eq:CS2_sj}
\end{eqnarray}
The  uncertainties mainly arise from the trigger efficiency and acceptance calculations. 

This value is compared to that calculated in the ``$k_T$ factorization" approach~\cite{barsnigzotnew}
with the unintegrated gluon density 
\cite{GluonPDF}:
\begin{eqnarray}
 \sigma_{\rm k_T}(J/\psi) = 23.0 \pm 8.5 ~\mbox{nb}.
\label{eq:kT_single}
\end{eqnarray}
In this calculation, the $J/\psi$ meson is produced either directly or through the radiative 
$\chi_{1(2)} \rightarrow J/\psi + \gamma$ process~\cite{barsnigzotnew}. 
The uncertainty is determined  by variations of 
the gluon PDF and scale variations by a factor of 2 with  respect to the default choice $\mu_R = \mu_F = \hat s/4$.

\begin{figure}[htb]
\includegraphics[width=0.5\textwidth,keepaspectratio=true]{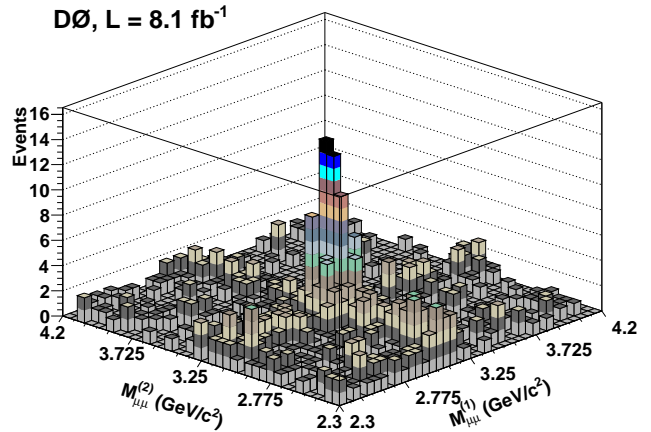} 
  \caption{(color online) Dimuon invariant mass distribution in data  for two muon pairs $M_{\mu\mu}^{(1)}$, $M_{\mu\mu}^{(2)}$ after the DJ selection criteria. }
  \label{fig:mass_2D}
\end{figure}

In total, 242 events remain after \DJP selection criteria and 902 events are found 
in the wider mass window $2.3 < M_{\mu \mu} < 4.2$ GeV.
Fig.~\ref{fig:mass_2D} shows the distribution of the two dimon masses ($M^{(1),(2)}_{\mu\mu}$) in these events. 

In analogy with the single $J/\psi$ event selection, we estimate the accidental, $J2\mu$  backgrounds and fraction
of prompt \DJP events. First, we reduce the non-prompt and background events
by requiring $c \tau < 0.03$ cm for both $J/\psi$ candidates, with about 94\% efficiency
for signal events (see Fig.~\ref{fig:single_jp}). This cut selects $N_{\rm d}=138$ events in data.  

\begin{figure}[h!]
~\\[-2mm]
\includegraphics[width=0.46\textwidth]{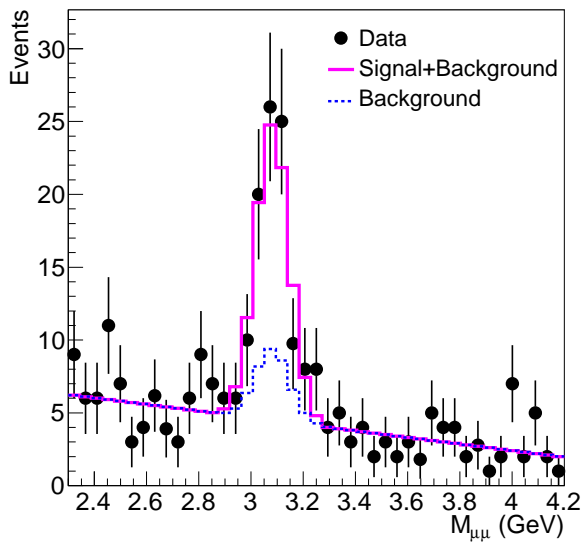}
\caption{(color online) Comparison of the signal and background contributions
to data projected on the axis of one muon pair $M_{\mu\mu}$ while events along the second pair
are integrated over the mass range $2.85-3.35$ GeV.}
\label{fig:bkg_fit}
\end{figure}

The signal and accidental background contributions are modelled using the product
\begin{equation} 
\begin{split}
F(M^{(1)}_{\mu \mu},M^{(2)}_{\mu \mu})  =&  (a_1 G^{(1)} + a_2 M^{(1)}_{\mu \mu} + a_3) \times \\ 
               & (a_4 G^{(2)} + a_5 M^{(2)}_{\mu \mu} + a_6),
\label{eq:2d_mass}
\end{split}
\end{equation}
where $a_{1(4)} G^{(1(2))}$ is a Gaussian function representing $J/\psi$ production, $a_{2(5)} M^{(1(2))}+a_{3(6)}$ 
is a linear function of the dimuon mass representing the accidental background, and $a_i$ are coefficients.  
%
To estimate the backgrounds in the selected data,  we perform a maximum likelihood fit to the data, 
in the two-dimensional (2D) ($M_{\mu\mu}^{(1)}$, $M_{\mu\mu}^{(2)}$) 
plane (see Fig.~\ref{fig:mass_2D}) using the expanded expression in Eq.~\ref{eq:2d_mass},
that contains a product of Gaussian functions for the signal \DJP mass peak 
while the background is represented by a plane (representing the accidental background) and 
a product of a Gaussian function and a line (for $J2\mu$ events).
We use the fitted parameters 
to estimate the  background in the signal window 
$2.85 < M_{\mu \mu} < 3.35$ GeV for both $J/\psi$ meson candidates 
and compute the fraction of the accidental+$J2\mu$ background events to be  $f_{\rm acc, J2\mu}=0.34 \pm 0.05$.
Figure \ref{fig:bkg_fit} shows a comparison of the summed signal and background contributions
to data projected on the axis of one muon pair $M_{\mu\mu}$ while events along the second pair
are integrated over the mass range $2.85-3.35$ GeV.

To estimate the fraction of the prompt double $J/\psi$ events, we use a template fit 
to the 2D $c \tau$ distribution in DJ data. 
The $c \tau$ template for double prompt 
mesons is obtained from the signal MC sample. 
The double non-prompt template is created from the $b \bar b$ MC sample, in which B hadron decays produce two $J/\psi$ mesons.
We also create a prompt+non-prompt template by randomly choosing  $c \tau$ values from the prompt and non-prompt templates.
Before fitting, the accidental and $J2\mu$ background is
subtracted from the data according to its fraction ($f_{\rm acc, J2\mu}$) with the uncertainty propagated 
into an uncertaity on the prompt fraction. The 2D $c \tau$ template for this backgound is built using 
data outside the signal mass window.
The prompt fraction of \DJP events in our selection is found to be
$f_{\rm prompt}=0.592 \pm 0.101$,
while the non-prompt and prompt+non-prompt events contribute $0.373 \pm 0.073$ and $0.035 \pm 0.073$, respectively.
The main source of systematic uncertainty for the prompt fraction is the template fitting, and the uncertainty related with  
the  subtraction of the accidental background  from the data.

We measure the acceptances, reconstruction, and selection efficiencies separately for double $J/\psi$ events on SP and DP samples using a mixture of 90\% color singlet and 10\% color octet samples, as predicted by NRQCD ~\cite{NRQCD}
for our kinematic selection criteria. The code for the predictions is implemented in the MC model DJpsiFDC ~\cite{DJFDC}.
We use {\sc pythia} for showering and fragmentation of the  $gg\to J/\psi J/\psi$ final state.
Products of the acceptances and the selection  efficiencies are found to be  $(A \varepsilon_{s})^{\rm{SP}}=0.109 \pm 0.002\mbox{(stat)} \pm 0.005\mbox{(syst)}$ for the SP and 
$(A\varepsilon_{s})^{\rm{DP}}=0.099 \pm 0.006\mbox{(stat)} \pm 0.005\mbox{(syst)}$ for the DP events, 
where the systematic uncertainties arise from uncertainties in 
 the modeling of  the $J/\psi$ kinematics, muon identification efficiencies and the possible non-zero $J/\psi$ polarization effects.

\begin{figure}[h!]
~\\[-2mm]
\includegraphics[width=0.49\textwidth,height=0.49\textwidth]{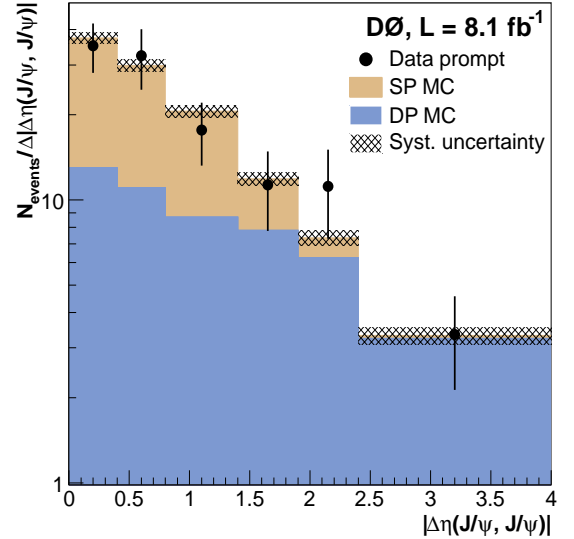}
~\\[-5mm]
\caption{(color online) 
The $|\Delta \eta (J/\psi, J/\psi)|$ distribution of  background subtracted double $J/\psi$ events
after all selection criteria. The distributions for the
SP and DP templates are shown normalized to their respective fitted fractions. 
The uncertainty band corresponds to the total systematic uncertainty on the sum of SP and DP events.}

\label{fig:djj_dp_fit}
\end{figure}

In this analysis, we measure the DJ production cross section for the DP and SP scatterings separately.
To discriminate between  the two mechanisms,
we exploit the distribution of the pseudorapidity difference 
between the two $J/\psi$ candidates, $|\Delta \eta (J/\psi,J/\psi)|$
which is stable against radiation and instrinsic parton $p_T$ effects~\cite{kom,barsnigzotnew}.
For the 
two $J/\psi$ mesons produced 
from two almost uncorrelated parton scatterings with smaller (on average) parton momentum fractions 
than in the SP  scattering, the $|\Delta \eta (J/\psi,J/\psi)|$ distribution is expected to be broader.
We use the DP and SP templates produced by MC
to obtain the DP and SP fractions from a maximum likelihood fit to the $|\Delta \eta (J/\psi,J/\psi)|$ distribution in \DJP data. 
Contributions from the accidental background, non-prompt and prompt+non-prompt
double $J/\psi$ events are subtracted from data.
The fit result is shown in Fig.~\ref{fig:djj_dp_fit}.
In the region $|\Delta \eta (J/\psi, J/\psi)|\gtrsim 2$, the data are dominated by DP events,
as predicted in Ref.~\cite{barsnigzotnew}.
A possible contribution from pseudo-diffractive gluon-gluon scattering 
should give a negligible contribution~\cite{barsnigzotnew}.
To estimate the systematic uncertainties of the DP and SP fractions, 
we vary the subtraction of accidental, non-prompt and prompt+non-prompt backgrounds within their uncertainties.
To conservatively estimate  systematic uncertainty related to the prompt+non-prompt background,
it is assumed to be either 100\% SP- or DP-like.
We also create a data-like DP template combining two $J/\psi$ meson candidates from 
two events randomly selected from the single $J/\psi$ data sample, 
emulating two independent scatterings each with a single $J/\psi$ final state.
This template is  corrected for the accidental and non-prompt backgrounds in data.
We extract the DP and SP fractions from the fit to the DJ data sample.
We find the fractions to be 
$f^{\rm DP}=0.42 \pm 0.12$ and $f^{\rm SP}=0.58 \pm 0.12$. 
These results are averaged over those obtained with the two SP ({\sc herwig++} and DJpsiFDC) 
and two DP ({\sc pythia} and data-like) models.
The main sources of the uncertainties on  DP (SP) fractions are  the background subtraction, 
18.4\% (13.4\%), the model dependence, 19.3\% (14\%), and the template fit, 7.1\% (6.3\%).
The uncertainty due to the model dependence is 
estimated by varying the DP  and SP models, and 
mainly caused by the difference between the two DP models.
Variation of the gluon PDF~\cite{GluonPDF} results in a small change 
of the DP and SP $|\Delta \eta (J/\psi,J/\psi)|$ templates and introduces a negligible uncertainty on the DP fraction.
We verify that we do not introduce a bias by determining the prompt, SP, and DP fractions in data by doing
two successive fits of the $c\tau$ and $|\Delta \eta (J/\psi, J/\psi)|$ distributions. 
For this purpose, we perform a simultaneous 2D fit for the non-prompt, SP, and DP fractions using templates
as functions of inclusive $c \tau$ and $|\Delta \eta (J/\psi, J/\psi)|$ to the data corrected for the accidental
and prompt+non-prompt backgrounds. The fractions of prompt DP and SP events determined by this procedure are in agreement
within uncertainties with the central result obtained by the two successive fits.

The fiducial prompt \DJP cross section is calculated according to 
\begin{eqnarray}
\sigma(J/\psi J/\psi) \!=\!\!\frac{N_{\rm{d}} f_{\rm prompt}\!(1-f_{\rm acc,J2\mu})}{L}\!\!\!\sum\limits_{i = {\rm DP,SP}}\frac{f^i}{(A\varepsilon_{\rm{s}})^i \!\varepsilon_{tr}^i},
 \label{eq:CS1}
\end{eqnarray}
where $N_{\rm{d}}$ is the number of data events in the \DJP selection, $f_{\rm prompt}$ is the fraction of prompt \DJP events, 
$f^i$ is the fraction of DP or SP events,
$\varepsilon_{\rm{tr}}^i$ is the trigger efficiency,  $(A\varepsilon_{s})^i$ is 
the product of acceptance and selection and reconstruction efficiency, and $L$ is the integrated luminosity.

Using the numbers presented above, we obtain
\begin{eqnarray}
 \sigma(J/\psi J/\psi)=129\pm11\mbox{(stat)}\pm37\mbox{(syst)}~\mbox{fb}. 
 \label{eq:CS2}
\end{eqnarray}
In the same way, we calculate the cross sections of DP and SP events individually 
\begin{eqnarray}
\sigma_{\rm{DP}}(J/\psi J/\psi)=59\pm6\mbox{(stat)}\pm22\mbox{(syst)}~\mbox{fb},
 \label{eq:CS2_djj_DP}
\end{eqnarray}
 \begin{eqnarray}
\sigma_{\rm{SP}}(J/\psi J/\psi)=70\pm6\mbox{(stat)}\pm22\mbox{(syst)}~\mbox{fb}.
\label{eq:CS2_djj_SP}
\end{eqnarray}

\noindent The prediction for the SP cross section made in the ``$k_T$ factorization'' approach~\cite{barsnigzotnew} is
\begin{eqnarray}
\sigma_{\rm k_T}(J/\psi J/\psi)  =55.1^{+28.5}_{-15.6} ({\rm PDF}) ^{+31.0} _{-17.0} ({\rm scale}) ~\mbox{fb}.
\label{eq:kT}
\end{eqnarray}
The choice of the gluon density as well as the renormalization and factorization scales are the same 
as for the prediction shown in Eq.~\ref{eq:kT_single}.

We also compare our $\sigma_{\rm{SP}}(J/\psi J/\psi)$ result to the SP prediction 
obtained with NRQCD at the leading order approximation in the strong coupling
~\cite{NRQCD} using renormalization and factorization 
scales of $\mu_R = \mu_F = ((\mbox{\ptjs})^2+m^2_c)^{1/2}$ and $m_c = 1.5$ GeV 
\begin{eqnarray}
\label{eq:nrqcd}
\sigma_{\rm NRQCD}^{\rm LO}(J/\psi J/\psi) = 51.9 ~\mbox{fb},
\end{eqnarray}
and NRQCD NLO predictions \cite{NRQCD_nlo}
\begin{eqnarray}
\label{eq:nrqcd_nlo}
\sigma_{\rm NRQCD}^{\rm NLO}(J/\psi J/\psi) = 90^{+180}_{-50} ~\mbox{fb}, 
\end{eqnarray}
where the uncertainty is due to the $\mu_R$ and $\mu_F$ scale variations by a factor two
as well as by the $c$-quark mass uncertainty $m_c = 1.5\pm0.1$ GeV.

The measured SP cross section is in agreement with the current predictions from NRQCD and 
``$k_T$ factorization''.
%

The DP production cross section predicted by the ``$k_T$ factorization'' approach 
according to Eq.~\ref{eq:s_eff},  
and using the fixed  effective cross section $\sigma^0_{\rm eff} = 15$ mb~\cite{barsnigzotnew}, is 
\begin{eqnarray}
\sigma_{\rm k_T}^{\rm DP}(J/\psi J/\psi) = 17.6 \pm 13.0 ~\mbox{fb}.
\label{eq:kTDP}
\end{eqnarray}

Additional contributions to the prompt DJ production may be caused by decays $\psi(2S)\to J/\psi+X$,
which are not taken into account in 
Eqs.~\ref{eq:kT} -- \ref{eq:kTDP}.
These contributions may increase the predicted DJ SP and DP cross sections by approximately $40\pm20\%$ \cite{talks}.


Using the measured cross sections of prompt single $J/\psi$ and DP production, we calculate the effective cross section, \sigeff~(see Eq.~\ref{eq:s_eff}). 
The main sources of systematic uncertainty in the \sigeff~measurement are trigger efficiency and the fraction of DP events.
By substituting the measured single $J/\psi$ and double $J/\psi$ DP cross sections 
(Eqs.~\ref{eq:CS2_sj} and \ref{eq:CS2_djj_DP}) into Eq.~\ref{eq:s_eff},
we obtain
\begin{eqnarray}
 \sigma_{\rm eff} = 4.8 \pm 0.5 \mbox{(stat)} \pm 2.5 \mbox{(syst)} ~\mbox{mb}.
 \label{eq:eff_sigma_0}
\end{eqnarray}

In conclusion, we have observed  double $J/\psi$ production at the Tevatron and measured its cross section. 
We show that this production is caused by single and double
parton scatterings. The measured SP cross section may indicate a need
for a higher gluon PDF at small parton momenta and small energy scale,
and higher order corrections to the theoretical predictions.
The measured \sigeff~agrees with the result reported by the AFS Collaboration ($\approx 5$ mb~\cite{afs2}),
and is in agreement with the \sigeff~ obtained by CDF~\cite{cdf1}
 in the 4-jet final state ($12.1^{+10.7}_{-5.4}$ mb).
However, it is lower than the result obtained by CDF~\cite{cdf2} ($14.5 \pm 1.7({\rm stat}) ^{+1.7} _{-2.3}({\rm syst})$)
and the D0~\cite{D0_2} result ($12.7 \pm 0.2({\rm stat}) \pm 1.3({\rm syst})$) in $\gamma + 3$-jet
events, and by ATLAS~\cite{atlas} ($15 \pm 3({\rm stat}) ^{+5} _{-3}({\rm syst})$) and by CMS~\cite{cms_w2j} ($20.7 \pm 0.8({\rm stat}) \pm 6.6 ({\rm syst})$) 
in the $W$+2-jet final state. 
We note that initial state in the DP double $J/\psi$ production is very similar to 4-jet production at low $p_T$ 
which is dominated by gluons, while $\gamma(W)$+jets events are produced in quark interactions, $q \bar{q}, qg$, and $q\bar{q}'$. 
The measured \sigeff~may indicate a smaller average distance between gluons than between quarks or between a quark and a gluon, in the transverse space.
This result is in a qualitative agreement with the pion cloud model predicting a smaller nucleon's 
average gluonic transverse size than that for singlet quarks~\cite{cloud_mpd}. 

%
We are grateful to the authors of the theoretical calculations, 
S.P.~Baranov, N.P.~Zotov, A.M.~Snigirev, C.-F.~Qiao, J.-P.~Lansberg, H.-S. Shao, and M.~Strikman
for providing predictions and for many useful discussions. 

We thank the staffs at Fermilab and collaborating institutions,
and acknowledge support from the
DOE and NSF (USA);
CEA and CNRS/IN2P3 (France);
MON, NRC KI and RFBR (Russia);
CNPq, FAPERJ, FAPESP and FUNDUNESP (Brazil);
DAE and DST (India);
Colciencias (Colombia);
CONACyT (Mexico);
NRF (Korea);
FOM (The Netherlands);
STFC and the Royal Society (United Kingdom);
MSMT and GACR (Czech Republic);
BMBF and DFG (Germany);
SFI (Ireland);
The Swedish Research Council (Sweden);
and
CAS and CNSF (China).
%



\begin{thebibliography}{99}
\bibitem{coord}
Pseudorapidity is defined as $\eta = -\ln[\tan(\theta/2)]$, where $\theta$ is the polar angle with
respect to the positive $z$ axis along the proton beam direction, while the azimuthal angle $\phi$ is defined
with respect to the $x$ axis pointing away from the center of the Tevatron ring. The $y$ axis points upwards.
\bibitem{afs2}
 In Ref.~\cite{AFS} the AFS collaboration does not report
   an uncertainty on \sigeff~but it is expected to be at
   least 30\% given the uncertainty quoted on the measured DP fraction.
\end{thebibliography}
\end{document}